\documentclass[12pt,cite]{article}
\input epsf.tex 
\def\beq{\begin{equation}} 
\def\eeq{\end{equation}} 
\def\eeq{\end{equation}} 
\def\bea{\begin{eqnarray}} 
\def\eea{\end{eqnarray}} 
\def\bq{\begin{quote}} 
\def\eq{\end{quote}}

\makeatletter 
\def\vereq#1#2{\lower3pt\vbox{\baselineskip1.5pt \lineskip1.5pt 
\ialign{$\m@th#1\hfill##\hfil$\crcr#2\crcr\sim\crcr}}} 
\makeatother

\title{Little Higgses and Turtles}

\author{
        David E. Kaplan,$^a$\,\thanks{\tt dkaplan@pha.jhu.edu}\ \
        Martin Schmaltz,$^b$\,\thanks{\tt schmaltz@bu.edu} \ \ 
        Witold Skiba$^c$\,\thanks{\tt witold.skiba@yale.edu} \ \ \\ \\ 
        \small \sl $^a$\ Department of Physics and Astronomy,
                Johns Hopkins University,\\  \small \sl 3400 N. Charles St., 
                Baltimore, MD  21218-2686\\
         \small \sl $^b$\ Department of Physics, Boston University, 
                Boston, MA  02215\\
                \small \sl $^c$\ Department of Physics, Yale University, 
                New Haven, CT  06520\\ \\ \\ 
       }

\begin{document} 
\baselineskip=17pt 
\pagestyle{plain}

\begin{titlepage} 
\vskip-.4in 
\maketitle 

\begin{abstract} 
\leftskip-.6in 
\rightskip-.6in 
\vskip.4in 

We present an ultra-violet extension of the ``simplest little Higgs'' model. 
The model marries the simplest little Higgs at low energies to two copies
of the ``littlest Higgs'' at higher energies.  The result is a weakly coupled
theory below 100 TeV with a naturally light Higgs.  The higher cutoff suppresses
the contributions of strongly coupled dynamics to dangerous operators such
as those which induce flavor changing neutral currents and CP violation. 
We briefly survey the distinctive phenomenology of the model.

\end{abstract} 
\thispagestyle{empty} 
\setcounter{page}{0} 
\end{titlepage}

\section{Introduction} 

The Standard Model (SM) is well supported by all high energy data.  Precision
tests match predictions which include one-loop quantum corrections.
These tests suggest that the SM is a valid description of Nature up to 
energies of several TeV with a Higgs mass which is less than
about 200 GeV\@.

On the other hand, the Standard Model is incomplete as quadratically
divergent quantum corrections to the Higgs mass destabilize the electroweak scale.
Naturalness of the SM with a light Higgs boson requires that the cutoff,
and therefore new physics, can not be significantly higher than about 1~TeV\@.
This implies an interesting conundrum:
naturalness wants new physics at a scale close to one TeV\@,
but the good fit of Standard Model predictions to the precision electroweak 
measurements severely constrains new physics up to several TeV\@.

Supersymmetry and little Higgs theories\cite{ACG2,Arkani-Hamed:2002qx,
Arkani-Hamed:2002qy,Gregoire:2002ra,Low:2002ws,Kaplan:2003uc,Chang:2003un,
Skiba:2003yf,Chang:2003zn,Cheng:2003ju} can solve this puzzle by
introducing new weakly coupled particles near 1 TeV which cancel
the problematic quadratic divergences and are hidden sufficiently well
from precision electroweak experiments~\cite{EW1,EW2}.
In little Higgs theories, this cancelation is ensured
by a set of approximate bosonic symmetries, and the cutoff can be raised
to about $4 \pi$ times 1~TeV\@. 

10~TeV is beyond the reach of LHC experiments, making direct
exploration of the physics above the cutoff unrealistic for the near future.
However, rare processes like baryon number violation
and flavor or CP violation indirectly probe significantly larger scales.
We are therefore motivated to explore whether Little Higgs theories
can be embedded in  UV extensions which go beyond 10~TeV
and are in agreement with the indirect constraints. Previous work in
this direction includes a study of flavor in generic strongly
coupled UV extensions  \cite{Chivukula:2002ww}, and the
construction of a strongly coupled UV completion with supersymmetry \cite{Katz:2003sn}.

In this paper we construct a {\it weakly coupled} UV extension
by embedding the nonlinear sigma model fields of the little Higgs theory
into a linear sigma model at the scale $f\sim\, $TeV.
The linear sigma model remains weakly coupled at 10~TeV.
However, since it contains fundamental scalars it suffers from a naturalness
problem at scales above 10~TeV. We solve this ``upstairs'' naturalness problem
by applying the little Higgs mechanism again. The  ``upstairs'' little
Higgs theory is strongly coupled at  $\Lambda\sim 100\, $TeV.
At the symmetry breaking scale of the upstairs theory, $F\sim 10\, $TeV\@,
we match onto the ``downstairs'' little Higgs model. At $f\sim1\, $TeV
the symmetries of the downstairs theory break, yielding a naturally light 
Standard Model Higgs. We refer to a little Higgs theory which is UV
extended by a little Higgs theory as a turtle, borrowing the image
from Hindu mythology as popularized by Hawking \cite{hawking}.

Concretely, the downstairs little Higgs is the ``simple gauge group'' $SU(3)$
model of Kaplan and Schmaltz \cite{Kaplan:2003uc,BS}, and the upstairs little Higgs theory
consists of two copies of the ``littlest Higgs''~\cite{Arkani-Hamed:2002qy},
generalized to $SU(7)/SO(7)$ to incorporate the larger $SU(3)$ gauge symmetry
of the downstairs theory.

To keep the theory natural up to the cutoff at 100 TeV one must eliminate
all large corrections to the Higgs boson mass. Specifically, we must eliminate
quadratic divergences up to two loops because
$\delta m_h^2 \sim \Lambda^2/(16 \pi^2)^2 \approx f^2$ is too large,
in a natural theory radiative corrections should be of the same order as the Higgs mass.
In addition, one loop logarithmic and one loop finite corrections
must be suppressed. This is because a one loop finite diagram could give
$\delta m_h^2 \sim \frac{F^2}{16 \pi^2} \approx f^2$ times gauge
or Yukawa couplings squared which is also too large.

Thus, the problem of UV completing a little Higgs theory with another little Higgs
theory cannot rely on eliminating quadratic divergences only.  In fact, this was
already accomplished in the original little Higgs model~\cite{ACG2}.  
The real key is to suppress the scale at which finite one-loop contributions
to the Higgs mass appear.  In particular, one must ensure that dimension-two
operators are sufficiently suppressed.
In our low-energy $SU(3)$ theory with two fundamentals, there are two such
operators, namely $|\phi_i |^2$ $(i=1,2)$ and $\phi_1^\dagger \phi_2$. 
The former is harmless because it does not break either
of the $SU(3)$ symmetries of
which the Higgs is an approximate Nambu-Goldstone boson; the latter breaks
the two $SU(3)$ symmetries to the diagonal and contains a direct Higgs mass.
Fortunately, the $\phi_1^\dagger \phi_2$ operator is not invariant under
a $Z_2$ symmetry under which $\phi_1$ is odd and $\phi_2$ is even (or vice-versa). 
Since the $Z_2$ parity symmetry is preserved by gauge interactions and the Yukawa couplings
radiative corrections do not generate the dangerous operator.
This is the main point of our paper.

In the following sections we describe our model from the top down. In section 2, we describe two
$SU(7)/SO(7)$ modules each of which generates an $SU(3)$ triplet little Higgs field $\phi_i$.
We discuss gauge and Yukawa interactions and prove that all large corrections to
the Higgs mass are eliminated. In Section 3 we describe the breaking of $SU(3)$ to $SU(2)$
which leads to an even lighter scalar doublet, the Higgs. In the final section we present
our conclusions, discuss some general features of the phenomenology and suggest future work.

\section{High energy little Higgs}
\label{sec:upstairs}

Our model is based on two copies of the nonlinear sigma model describing $SU(7)/SO(7)$
coset space. Each of the sigma models is a straightforward extension of the 
littlest Higgs, Ref.~\cite{Arkani-Hamed:2002qy}, designed to incorporate
a larger unbroken gauge symmetry --- $SU(3)$ instead of $SU(2)$. We review the
model and introduce our notation next.

\begin{figure}
\begin{picture}(400,200)(0,0)
\thicklines
\put(150,200){\oval(80,40)}
\put(310,200){\oval(80,40)}
\put(230,120){\oval(80,40)}
\put(230,40){\oval(80,40)}
\put(230,95){\vector(0,-1){30}}
\put(150,175){\vector(1,-1){40}}
\put(310,175){\vector(-1,-1){40}}

\thinlines
\put(116,196){$SU(7)/SO(7)$}
\put(276,196){$SU(7)/SO(7)$}
\put(10,200){upstairs}
\put(40,160){$F$}
\put(10,120){downstairs}
\put(40,80){$f$}
\put(10,40){Standard Model}
\put(195,116){$SU(3)\times U(1)$}
\put(180,150){$\phi_1$}
\put(267,150){$\phi_2$}
\put(195,36){Higgs doublet}
\end{picture}
\caption{Overall structure of the model. Two upstairs $SU(7)/SO(7)$ littlest Higgs models
each yield one $SU(3)$ triplet scalar $\phi$ with appropriate interactions
for the downstairs $SU(3)$ ``simple gauge group'' little Higgs model.}
\end{figure}
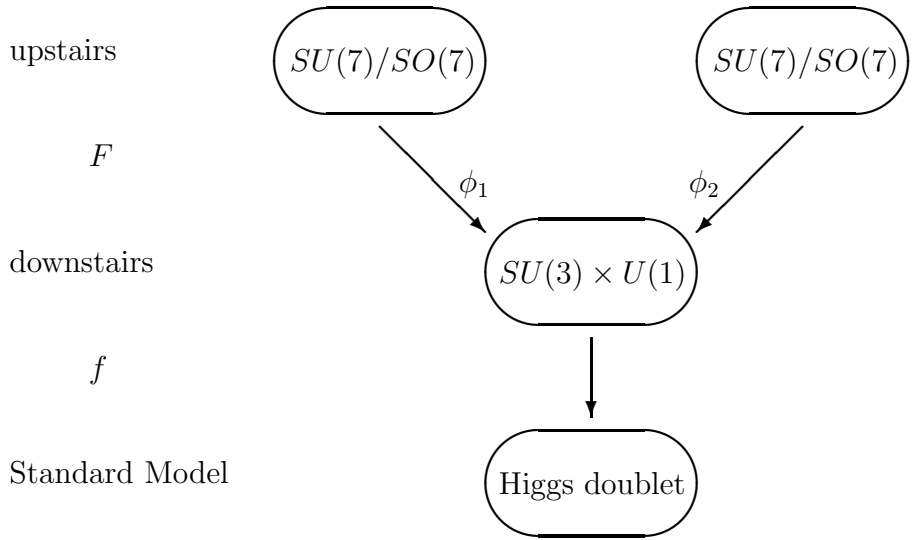

\subsection{Symmetries and gauge interactions}

Let us call the fields transforming linearly under two $SU(7)$ global symmetries
$\Sigma_i$, where $i=1,2$. Under an $SU(7)_i$ transformation $U_i$, the $\Sigma$ fields
transform as $\Sigma_i \rightarrow U_i \Sigma_i U_i^T$.  An $SU(3) \times SU(3) \times U(1)$
gauge symmetry is embedded in each $SU(7)$ in analogy with the gauge group embedding
in the littlest Higgs model~\cite{Arkani-Hamed:2002qy}. Below the scale $F$ of
$SU(7)_i\rightarrow SO(7)_i$ symmetry breaking, there will be an unbroken $SU(3)\times U(1)$
gauge symmetry and a light scalar triplet of the $SU(3)$. The breaking scales of the two sigma
models do not need to be identical, however they are chosen to be comparable.

In addition to color and a $U(1)_X$ symmetry which will be described below,
we gauge three $SU(3)$ symmetries.
The $SU(3)$ gauge symmetry generators are embedded in 
the two global $SU(7)$ symmetry groups in following way
\begin{equation}
\label{eq:SU3generators}
T_{u,d}^a= \left( \begin{array}{c|c|c} t_1^a & 0 & 0  \\ \hline 0 & 0 & 0 \\ \hline 0 & 0 & -(t^a_2)^*  \end{array}
 \right) \subset U_1  \;\; {\rm and} \; \; 
 T_{u,d}^a= \left( \begin{array}{c|c|c} t_1^a & 0 & 0  \\ \hline 0 & 0 & 0 \\ \hline 0 & 0 & -(t^a_3)^*  \end{array}
 \right) \subset U_2,
\end{equation} 
where $t^a$ are the $SU(3)$ Gell-Mann matrices. $T^a_u$ ($T_d^a$) refer to the generators 
embedded in the upper (lower) corner of the $SU(7)$ space. Our shorthand notation for
the 7 by 7 matrices is that the first and the last column or row have three components,
while the middle column and row have one component. Due to this embedding there is
only one unbroken $SU(3)$ at energies below $F$. We will refer to
this unbroken group that is a linear combination of all three $SU(3)$'s as $SU(3)_L$
because it contains the weak $SU(2)_L$ as  a subgroup.
There are $27$ Goldstone bosons in each $SU(7)\rightarrow SO(7)$  sigma
model. Under the unbroken $SU(3)_L$ these decompose as real ${\bf 8}$ and ${\bf 1}$
as well as complex ${\bf 6}$ and ${\bf 3}$. The octet representation is eaten in each
of the sigma model due to the breaking of $SU(3) \times SU(3)$ to the diagonal
$SU(3)_L$. 

The singlets would also be eaten if we gauged $U(1)^3$ instead of a single
$U(1)_X$ in a similar way to the gauging of $SU(3)^3$. The singlet
fields can be much lighter than the electroweak scale and still escape detection. They can
obtain masses by including for each $\Sigma$ a small explicit symmetry breaking term,
for example $\alpha\,( {\rm det}_{3\times3}(\Sigma_1) + {\rm h.c.})$ and an identical term
for $\Sigma_2$. The determinant ${\rm det}_{3\times3}$ is computed over a 3 by 3
dimensional subset of the indices that span 1st through 3rd rows of $\Sigma$ and
5th through 7th column. This operator gives masses to the singlet field as well as
the ${\bf 6}$ and ${\bf 3}$. If the coefficients of these operators are small then their
only relevant contributions are to singlet masses,  other fields obtain larger masses from elsewhere.

The $U(1)_X$ is embedded identically in both sigma models.
Its generator is
\begin{equation}
\label{eq:Xgenerator}
 X= \left( \begin{array}{c|c|c} \frac{1}{3} & 0 & 0  \\ \hline
             0 & 0 & 0 \\ \hline 0 & 0 & -\frac{1}{3}    \end{array} \right).
 \end{equation}
Given the symmetry properties of the $\Sigma_i$ fields and the gauge generators, the
covariant derivatives of these fields are
\begin{eqnarray}
 D_\mu \Sigma_1 &=& \partial_\mu \Sigma_1 + i g_1 T^a_u A^{a1}_\mu \Sigma_1 +
     i g_2 T^a_d  A^{a2}_\mu \Sigma_1 + i g_X X X_\mu \Sigma_1 \nonumber \\ &&
     + i g_1 \Sigma_1 (T^a_u)^T  A^{a1}_\mu +
     i g_2 \Sigma_1  (T^a_d)^T  A^{a2}_\mu+ i g_X \Sigma_1 X^T X_\mu,  \\
   D_\mu \Sigma_2 &=& \partial_\mu \Sigma_2 + i g_1 T^a_u A^{a1}_\mu \Sigma_2 +
     i g_3 T^a_d  A^{a3}_\mu \Sigma_2 + i g_X X X_\mu \Sigma_2 \nonumber \\  &&
     +i g_1 \Sigma_2 (T^a_u)^T  A^{a1}_\mu +
     i g_3 \Sigma_2  (T^a_d)^T  A^{a3}_\mu+ i g_X \Sigma_2 X^T X_\mu. 
\end{eqnarray}

In the unitary gauge which takes into account $SU(3)^3$ breaking (i.e. where we have gauged 
away fields which are eaten by the heavy gauge bosons), the 
$\Sigma_i$ fields can be parameterized as follows:
\begin{equation}
   \Sigma_i= e^{i \Pi_i/F_i} \Sigma_0, \; {\rm where} \; 
   \Pi_i=\left( \begin{array}{c|c|c} 0 & \phi^*_i  & S^*_i
                     \\ \hline \phi^T_i & 0 & \phi^\dagger_i \\ \hline S_i & \phi_i & 0 \end{array} \right)
               \; {\rm and} \;   \Sigma_0=\left( \begin{array}{c|c|c} 0 &0 & 1
                     \\ \hline 0 & 1 & 0 \\ \hline 1 & 0 & 0 \end{array} \right).
\end{equation}
Since the broken generators obey $B_a \Sigma_0 = \Sigma_0 B_a^T$, one can also
write $\Sigma_i=\Sigma_0 e^{i \Pi_i^T/f}$. Here, $\phi_i$ are complex
triplets and $S_i$ are complex symmetric tensors of $SU(3)$. We did not include
the singlet fields since they do not play any role in the
following discussion. Canonically normalized kinetic terms for these fields
are obtained from expanding out
\begin{equation}
  {\mathcal L}_{kin}= \frac{F_1^2}{4}  {\rm tr}\left[ (D^\mu \Sigma_1)^* D_\mu \Sigma_1\right]
                                 + \frac{F_2^2}{4}  {\rm tr}\left[ (D^\mu \Sigma_2)^* D_\mu \Sigma_2\right].
\end{equation}

We now turn to a discussion of the potential which is generated for the $SU(3)$ triplet
fields $\phi_i$. We wish the potential to be of the form
\begin{equation}
{V}\sim (\phi_1^\dagger \phi_1 -f_1^2)^2 + (\phi_2^\dagger \phi_2 -f_2^2)^2
+ B \, \phi_1^\dagger \phi_2 + \lambda (\phi_1^\dagger \phi_2)^2 + ...
\label{eq:phipot}
\end{equation}
The first two terms are necessary to ensure the correct symmetry breaking in the
downstairs little Higgs theory whereas the last two terms contain a mass term
and the quartic coupling for the Standard Model Higgs. In order for electroweak
symmetry breaking to be natural and for the Higgs mass to be above
the LEP2 bound, we need $B$ to be of order the weak scale squared
and $\lambda \geq 0.1$.

The first and second terms each contain only one of the two $\phi_i$, they are
generated from gauge and Yukawa interactions exactly as in the littlest Higgs
model \cite{Arkani-Hamed:2002qy}. The fourth term involves fields from both
$\Sigma_1$ and $\Sigma_2$. It is easy to see that the leading power
divergent diagrams cannot generate this term because there is no direct
coupling between $\Sigma_1$ and $\Sigma_2$. Subleading diagrams give a coefficient 
of the desired size. The third term, if generated at all, would be dangerous.
This is because naive power counting of finite diagrams with loop momenta
of order $F$ give a coefficient of order $F^2/16 \pi^2 \sim f^2$ which is too large.
However, there are global $Z_2$ symmetries which are respected by the gauge
interactions and the Yukawa couplings which forbid this term. Therefore 
the $B$ term is not radiatively
generated and must be included in the tree level Lagrangian.

In the following we discuss the different contributions to this potential in detail.
The gauge interactions explicitly break the $SU(7)$ symmetries associated with the
two sigma models, therefore gauge loops contribute to the potential for the
pseudo-Goldstone bosons. First, there are quadratically divergent contributions,
which are similar to the ones found in the littlest Higgs model~\cite{Arkani-Hamed:2002qy}:
\begin{eqnarray}
 {\mathcal L}_{\rm quad} &=&-c \sum_{a=1}^8  \left[ g_1^2 F_1^2 {\rm tr}(T_u^a \Sigma_1 T^*_{ua} \Sigma^*_1)
  +g_1^2 F_2^2 {\rm tr}(T_u^a \Sigma_2 T^*_{ua} \Sigma^*_2) \right. \nonumber \\
  && \left. + g_2^2 F_1^2  {\rm tr}(T_d^a \Sigma_1 T^*_{da} \Sigma^*_i)
  + g_3^2 F_2^2{\rm tr}(T_d^a \Sigma_2 T^*_{da} \Sigma^*_2) \right], \label{eq:gaugequad}
   \end{eqnarray}
where $T_u$ and $T_d$ refer to the $SU(3)$ generators embedded, respectively, in the
upper and lower corners of the $SU(7)$ transformations; $c$ is an ${\mathcal O}(1)$  constant determined by the dynamics at the cutoff scale.  Because the $\pi^\pm$--$\pi^0$ mass
difference is positive in QCD we assume that $c>0$. Expanding
Eq.~(\ref{eq:gaugequad}) in terms of the pseudo-Goldstone fields yields
\begin{eqnarray}
 {\mathcal L}_{\rm quad} &=& -\frac{4 c}{3} \left[g_1^2 F_1^2 |S_1^{mn} -
     \frac{i}{2 F_1} \phi_1^m \phi_1^n |^2
    + g_1^2 F_2^2 |S_2^{mn} - \frac{i}{2F_2} \phi_2^m \phi_2^n |^2 \right.\nonumber \\
&&  \left.  +g_2^2 F_1^2 |S_1^{mn} + \frac{i}{2F_1} \phi_1^m \phi_1^n |^2
  + g_3^2 F_2^2 |S_2^{mn} + \frac{i}{2 F_2} \phi_2^m \phi_2^n |^2 \right].
  \end{eqnarray}
Integrating out the massive $S_i$ fields yields a quartic potential for $\phi_i$'s
\begin{equation}
\label{eq:Vgauge}
 V= \frac{4 c}{3} \left[ \frac{g_1^2 g_2^2}{g_1^2+g_2^2} \, (\phi_1^\dagger \phi_1)^2 + 
       \frac{g_1^2 g_3^2}{g_1^2+g_3^2}\, (\phi_2^\dagger \phi_2)^2 \right].
\end{equation}

The $U(1)_X$ interactions also induce quadratically divergent contributions
to the masses of scalars which are small enough because of the small $U(1)_X$ gauge
coupling. The quadratically divergent contribution to $\Sigma_1$
is
\begin{equation}
\label{eq:VgaugeX}
{\mathcal L}_X= c g^2_X F_1^2 {\rm tr}(X \Sigma_1 X \Sigma^*_1)  
 =c g_X^2 F_1^2  2 ( 2 |S_1|^2 + |\phi_1|^2 +{\rm higher\ order})
\end{equation}
and there is an analogous contribution to $\Sigma_2$.

As we mentioned in the Introduction, we need to make sure there are no corrections
to the Higgs mass that are larger than $f/(4 \pi)$.
Loops generate $|\phi_i|^2$ mass terms of order $\frac{F^2}{16 \pi^2} \sim f^2$.
Such terms do not contribute to the Higgs mass, instead they induce vevs for $\phi_i$
if the dominant contribution is negative (as it will be from the top sector).
The Higgs mass is protected by two independent $SU(3)$ symmetries associated
with the triplets $\phi_i$,  see Ref.~\cite{Schmaltz:2002wx}. 
Neglecting the $SU(3)_L$ gauge interactions, each triplet has a separate
$SU(3)_i$ global symmetry. Potentially dangerous mass terms must break
the separate global symmetries that is they must be proportional to $\phi_2^\dagger \phi_1$.
It is easy to see that such terms are not generated above scale $f$ by the following
symmetry argument. 

The gauge interactions respect two separate $Z_2$ symmetries under
which the $\Sigma_i$ fields transform independently as
$\Sigma_i \rightarrow R_i \Sigma_i R_i$, where
\begin{equation}
 \label{eq:parity}
    R_1=R_2= \left( 
     \begin{array}{c|c|c} 1 & 0 & 0  \\ \hline 0 & -1 & 0 \\ \hline 0 & 0 & 1 \end{array} \right)
 \end{equation}
The gauge generators, Eqs.~(\ref{eq:SU3generators}) and (\ref{eq:Xgenerator}), commute with
the $R_i$. The action of $R_i$ on the pseudo-Goldstone bosons is such that
$\phi_i$ are odd and all other fields are even. Thus terms with an odd number of $\phi_1$'s or 
$\phi_2$'s, like the B term $\phi_2^\dagger \phi_1$, cannot be generated
 by interactions above scale $f$.  This is crucial because this B term is the one term that breaks 
 the two global $SU(3)$ symmetries and gives the Higgs a mass. 
Quartic terms which break the $SU(3)'s$,\ like $|\phi_2^\dagger \phi_1|^2$, are generated, 
but are suppressed by a dimensionless loop factor and generate
a sufficiently small contribution to the Higgs mass $\langle \phi \rangle^2/16\pi^2 \sim v^2$. 
The $\phi$ vev spontaneously breaks the parity symmetry and produces
the B term in Eq.~(\ref{eq:phipot}).
As we will show, the Yukawa couplings also preserve the $R_i$ symmetries.
Therefore loop contributions from the fermion sector are harmless as well.

\subsection{Yukawa couplings}
\label{sec:Yukawa}
Let us describe the third generation of fermions first.
Because the weak $SU(2)_L$ group is enlarged to $SU(3)_L$, all electroweak doublets
need to be in the triplet representation of $SU(3)$. In order to avoid the anomalies
associated with the $SU(3)_L$, we embed the first two families of quarks differently from
the third family. We will discuss such an anomaly-free embedding afterwards.

For the third generation of quarks, we consider the following fermions:
six singlets $\eta_i$, $\zeta_i$, $\chi_i$,
and 7  $SU(3)$ triplets $Q_i$, $P_i$, $P_i^c$, $Q^c$. 
In all cases, the index $i$ runs from 1 to 2 and corresponds to the two $SU(7)/SO(7)$
sigma models. Before the breaking of $SU(3)^3$ symmetry
to the diagonal, the triplets $Q_i$, $Q^c$ and $P_i$ transform under the common $SU(3)$,
while $P^c_1$ and $P^c_2$ transform under the remaining two $SU(3)$. 
We add the following couplings
\begin{eqnarray}
{\mathcal L}_{top}&=&
\lambda_1 \left( \begin{array}{c} Q_1 \\ \eta_1 \\ P^c_1 \end{array} \right)^T
     \Sigma_1 
   \left( \begin{array}{c} P_1 \\ \chi_1 \\ 0 \end{array} \right)
   +\lambda_2 \left( \begin{array}{c} Q_2 \\ \eta_2 \\ P^c_2 \end{array} \right)^T
     \Sigma_2 
   \left( \begin{array}{c} P_2 \\ \chi_2 \\ 0 \end{array} \right)  \nonumber \\
   && + \lambda_3 F Q^c (Q_1+Q_2) + \lambda_4 F (\eta_1\zeta_1 + \eta_2 \zeta_2) +{\rm h.c.}
    \label{eq:Ltop}
   \end{eqnarray}
This is just one possible example of Yukawa couplings. It is not difficult to write
couplings, in which only complete $SU(7)$ fermion multiplets couple to $\Sigma_i$,
but doing so requires introducing more fields. Such complete $SU(7)$ multiplets
could potentially make it easier to UV extend the model even further.

Most of the fermions obtain masses of order $F$ from the fermion Lagrangian above.
The fermions which are light compared to $F$ comprise a triplet and two singlets. 
The triplet $t_L$ is a linear combination of $Q_1$ and $Q_2$
and the light singlets $\chi_{1,R}^c$ and $\chi_{2,R}^c$ are linear combinations of
$\chi_i$ and $\zeta_i$. The Yukawa couplings for the top quark and
its partner that follow from Eq.~(\ref{eq:Ltop}) are then 
\begin{eqnarray}
{\mathcal L}_{top}= t_L \phi_1 \chi_{1,R}^c + t_L \phi_2 \chi_{2,R}^c. 
   \end{eqnarray}

Let us examine radiative corrections induced by the fermions with momenta 
larger than $F$. There are no quadratic divergences since each $\Sigma$ field
couples to a complete $SU(7)$ multiplet.  Note that the fermion Lagrangian
is invariant under the $Z_2$ operators introduced in Eq.~(\ref{eq:parity}), where
all singlet fields $\eta_i$, $\zeta_i$, $\chi_i$ are odd
under $R_i$.  All $SU(3)$ fermion triplets are even under both parity transformations.
The invariance under the $R_i$ transformation insures that loops involving
the top quark and its heavy partners do not generate dangerous terms.

There are no log-divergent diagrams that involve two different sigma 
fields simultaneously. The largest important radiative corrections are the
logarithmic divergences involving  one sigma field. Such log-divergent corrections
give negative masses squared for the scalars $\phi_i$, which triggers the independent
breaking of $SU(3)_L$ to $SU(2)_L$ by each of the two $\phi_i$'s. 
The coset space of the downstairs little Higgs, $(SU(3)/SU(2))^2$,
is generated as a result of this radiative symmetry breaking.  The log-divergent diagrams
yield the potential for $\Sigma$
\begin{equation}
  V_f=\frac{-3 }{16\pi^2} \lambda_1^2 \log(\Lambda/F) \, {\rm tr}(M_f^2 \Sigma_1 P^u_4 \Sigma_1^*)  
\end{equation}
where  $P^u_4$ is the projection
operator that picks out the upper four components of  the $\Sigma$
\begin{equation}
P^u_4= \left( \begin{array}{c|c|c} 1 & 0 & 0  \\ \hline 0 & 1 & 0 \\ \hline 0 & 0 & 0  \end{array}
 \right),
 \end{equation}
while $M_f^2$ is matrix of the explicit mass terms
\begin{equation}
 M_f^2= \left( \begin{array}{c|c|c} \lambda_3^2 F^2 & 0 & 0  \\ \hline 0 
    & \lambda_4^2 F^2& 0 \\ \hline 0 & 0 & 0  \end{array} \right).
 \end{equation}
Expanding $V_f$ in terms of the component fields up to second order gives
\begin{equation}
\label{eq:Vfermions}
  V_f=\frac{-3 }{16\pi^2} F^2 \log(\Lambda/F) \lambda_1^2
    \left[ (\lambda_3^2 -\lambda_4^2) |\phi_1|^2 +
    \lambda_3^2 |S_1|^2 \right].
 \end{equation}
We assume that $\lambda_3>\lambda_4$ in order to trigger the breaking of $SU(3)$.
Similar expression holds for the log-divergent contributions associated
with the fermions that couple to $\Sigma_2$.
It is easy to understand why the coefficient of the $\phi_i$ mass terms are proportional
to $\lambda_3^2-\lambda_4^2$. If $\lambda_3=\lambda_4$ the fermion Lagrangian,
Eq.~(\ref{eq:Ltop}), has an approximate global
$SU(4)_1\times SU(4)_2$ symmetry, where $SU(4)_i$ acts
on the first four indices of $\Sigma_i$. One can assign 
$SU(4)_i$ transformations to the fermions that couple to the $\Sigma_i$
directly, and the mass terms do not break this symmetry when $\lambda_3=\lambda_4$.
The $SU(4)_i$ symmetry protects $\phi_i$ from having a  mass term.
Actually, the $SU(4)_1\times SU(4)_2$ symmetry is explicitly broken by the 
$Q^c (Q_1+Q_2)$ term in Eq.~(\ref{eq:Ltop}). 
The argument we just gave only applies to the log-divergent
terms because the explicit soft breaking term shows up in finite contributions.

Two-loop interactions could generate terms with coefficients of order $f^2 \sim \Lambda^2 / 
(16\pi^2)^2$ if such interactions are quadratically divergent.  However, the only dimension-two 
operators that could be generated by quadratically divergent loops are $|\phi_i |^2$, $i=1,2$.  
The only coupling which breaks the parity symmetry is
the dimensionful $B$ term and thus there 
are no quadratic divergent contributions to the 
$\phi_1^\dagger \phi_2$ operator at {\it any} loop order!

To complete the discussion of the third generation quarks, we comment on the bottom
Yukawa coupling. Our theory has a cutoff of about 100~TeV, so one should worry about 
a large one-loop quadratic divergence from the bottom Yukawa coupling.
We can see this divergence in the theory below the scale $F$ where
the bottom Yukawa coupling involves both $\phi_i$ fields. The bottom Yukawa is
\begin{equation}
  {\lambda_b\over f} \,\epsilon^{mnp}  t_{L,m} \phi_{1n}^* \phi_{2p}^* \,b_R.
\end{equation}
A fermion loop involving this coupling twice generates the operator
$|\phi_1^\dagger \phi_2|^2$ with quadratically divergent coefficient
$(\lambda_b/f)^2 \Lambda^2/(16 \pi^2) \sim \lambda_b^2 16 \pi^2$ which is
too large. We can get rid of this one loop quadratic divergence by coupling fields
to the $\Sigma_i$'s one at a time:
\begin{equation}
 {\mathcal L}_b=\hat{\lambda}_1 \epsilon^{mnp} t_{L,m} (\Sigma_1)_{4n} R_p + F R R^c + 
           \hat{\lambda}_2 (R^c)^m (\Sigma_2)_{4m} b_R.
\end{equation}
Integrating out the massive vector-like pair $R$ and $R^c$ produces the desired
coupling. The couplings $\hat{\lambda}_i$ induce quadratic divergences
contributing to $|\phi_i|^2$ mass terms because
the corresponding Yukawa couplings do not respect $SU(4)$ symmetries.
But the diagram which gives the operator $|\phi_1^\dagger \phi_2|^2$ is now finite,
with a coefficient $\sim   (\hat{\lambda}_1 \hat{\lambda}_2)^2/(16 \pi^2)$.

We now present a complete set of fermions that corresponds to the Standard Model with
three families at low energies. In general, in a spontaneously
broken gauge theory anomalies can be cancelled by spectator fermions with masses
which are at most $4\pi $ times the symmetry breaking vacuum expectation value.
This implies that $SU(3)_L \times U(1)_X$ anomalies must be canceled
within our effective theory whereas  the anomalies of the full
$SU(3)^3 \times U(1)_X$ gauge theory
may be canceled by unspecified fields at the cutoff $\Lambda$.
 In order to cancel the $SU(3)_L \times U(1)_X$ anomalies we assign different
quantum numbers to the first two generations of quarks than to the third generation
(see for example \cite{Pisano:ee}).

We now list the $(SU(3)_c,SU(3)_L)_{U(1)_X}$ quantum
numbers below scale $F$. 
The third generation of quarks consists of the left-handed triplet
$({\bf 3},\overline{{\bf  3}})_\frac{1}{3}$, 
two conjugate fields $(\overline{{\bf 3}},{\bf 1})_{-\frac{2}{3}}$ corresponding
to the right-handed top and its heavy partner and the conjugate right-handed bottom
quark in the $(\overline{{\bf 3}},{\bf 1})_{\frac{1}{3}}$.
The first and second generations of quarks are identical to each
other, and consist of the left-handed triplets  $({\bf 3},{\bf 3})_0$, conjugate
up quarks  $(\overline{{\bf 3}},{\bf 1})_{-\frac{2}{3}}$, conjugate down quarks and their heavy
partners in  the $(\overline{{\bf 3}},{\bf 1})_\frac{1}{3}$. Finally, all three generations
of leptons are treated on the same footing. Each lepton generation consists
of a triplet $({\bf 1},\overline{{\bf 3}})_{-\frac{1}{3}}$, a conjugate of the
right-handed charged lepton $({\bf 1},{\bf 1})_1$
and a conjugate of the heavy neutrino in the $({\bf 1},{\bf 1})_0$ representation.
It is straightforward to check that  the $SU(3)_L$ and $U(1)_X$ anomalies
as well as the mixed anomalies $(SU(3)_L)^2 U(1)_X$ 
and $(SU(3)_c)^2 U(1)_X$ all vanish. 

We have already shown how to incorporate the couplings of the third generation
of quarks above scale $F$. Below $F$, these couplings reduce to
$\phi_1^m t_{L,m} \chi_{1,R}^c + \phi_2^m t_{L,m} \chi_{2,R}^c+
{\lambda_b\over f} \,\epsilon^{mnp}  t_{L,m} \phi_{1n}^* \phi_{2p}^* \,b_R$, where we 
list only the $SU(3)_L$ contractions explicitly and $\phi_i$ are the scalar triplets
with the $U(1)_X$ charges of $\frac{1}{3}$.  All the remaining quarks
and leptons have small Yukawa couplings in the Standard Model,
thus they can be incorporated without regard to divergences 
such couplings may induce.  For completeness, we list the required couplings below scale 
$F$ for the first two generations of quarks and all generations of leptons. 
The up quarks get mass from the $\epsilon_{mno} \phi^m_1 \phi_2^n q_L^o u_R^c$ operator.
The down quarks and their partners get their masses from $ q_L^m  \phi^*_{1,m} d_R^c +
q_L^m \phi_{2,m}^* \chi_{d,R}^c$. The couplings for the charged leptons and the heavy neutrinos
are of the form $\epsilon^{mno}  \phi^*_{1,m} \phi^*_{2,n}  E_{L,o} e_R^c + 
E_{L,m} \phi_2^m \chi_{\nu,R}^c$, where $\chi_R$ indicate the heavy right-handed partners. 
Neutrino masses and mixings can be accommodated by adding the operators
$(E_L \phi_1)^2/\Lambda + (E_L \phi_2)^2/\Lambda$ with small
coefficients. Promoting these Yukawa couplings into the full theory is completely straightforward
and can be done, for example, by replacing $\phi_1^m$ with $(\Sigma_1)^{4m}$
and $\phi_2^m$ with $(\Sigma_2)^{4m}$ in the full theory. To the lowest order in
fields, $(\Sigma_i)^{4m}=\phi_i^m$.

 \section{Low energy little Higgs}
 \label{sec:downstairs} 

 Having discussed the interactions generated in the effective
 theory between scales $F$ and $\Lambda$ we now turn to the description of
 our model below $F$. Let us reiterate the field content below scale $F$.
 The gauge group is $SU(3)_L\times U(1)_X$.  In addition to the gauge bosons,
 there are two $SU(3)_L$ scalar triplets $\phi_1$ and $\phi_2$.  Both scalar triplets
 transform linearly under $SU(3)_L$ that is each has six real degrees of freedom
 and have charge $\frac{1}{3}$ under $U(1)_X$. At scale $F$, the leading interactions
in the scalar  potential are described by Eq.~(\ref{eq:phipot}).
  
  The fermions are either  $SU(3)_L$ singlets or in three-dimensional
 representations of $SU(3)_L$. We listed the charge assignment of all fermions
 under $SU(3)_L\times U(1)_X$ at the end of Sec.~\ref{sec:Yukawa}.
 The only numerically important Yukawa couplings are the ones corresponding
 to the top quark and its heavy partner. 
 These are $\lambda_{t,1}\, \phi_1 t_{L} t_R^c +  \lambda_{t,2}\, \phi_2 t_{L} \chi_{t,R}^c$,
 where $t_L$ is in the anti-fundamental representation of $SU(3)_L$, while $t_R^c$
 and $\chi_{t,R}^c$ are singlets under $SU(3)_L$.

 Our downstairs little Higgs theory is almost identical to the little Higgs model presented
 in Ref.~\cite{BS}. The only difference is that the scalars in our theory correspond
 to linear sigma models, while in Ref.~\cite{BS} the scalars correspond to nonlinear
 sigma models with each field having one fewer degree of freedom. 
 However, the analysis of the models is parallel in the two cases so we
 will not repeat it here. We comment on the radial components of the $\phi_i$'s only.
  We divide the scalar degrees of freedom into the radial components
 and nonlinear multiplets as follows 
 \begin{equation}
  \phi_i=e^{\pi_i^a t^a/f_i} \left(\begin{array}{c} 0 \\ 0 \\ f_i + r_i \end{array} \right),
\end{equation}
where $t^a$ are the Gell-Mann matrices corresponding to the
broken $SU(3)$ generators, $a=4,\ldots,8$. 

The scalar potential, Eq.~(\ref{eq:phipot}),
receives contributions from above and below $F$. 
The potential determines the vevs for $\phi_i$ and
the masses for the radial excitations $r_i$.
Since both the quartic and quadratic terms for the $\phi_i$ receive divergent 
contributions we cannot calculate the potential exactly.
The magnitudes of these contributions have been discussed before. 
 We therefore treat the masses and the coupling constants $B$ and $\lambda$
 as free parameters.
 
 One can ask if it is natural for the vevs of $\phi_1$ and $\phi_2$ to point
 in the same direction. First, the Coleman-Weinberg effective potential from
 the gauge loops gives negative $|\phi_1^\dagger \phi_2|^2$, which favors
 proper vev alignment.  After computing the full potential one still needs to check that
  there are no flat directions or tachyons. Indeed, this is what happens.

 When the triplets $\phi_i$ get vevs, the $SU(3)_L \times U(1)_X$ is broken to
 the Standard Model $SU(2)_L\times U(1)_Y$ gauge group as described in
 Refs.~\cite{Kaplan:2003uc,BS}. The remaining light scalars comprise a complex 
 doublet of $SU(2)_L$, the Higgs, and one real scalar. 
 Radiative corrections between the scales $f$ and $v$ generate additional
 contributions to the quartic Higgs potential and the Higgs mass.
 The radiatively generated potential when combined with the tree-level B term
 $-\mu^2 \Phi_1^\dagger \Phi_2 + {\rm h.c.}$ leads to electroweak symmetry breaking.
 In order for the weak scale to come out correctly, the parameter $\mu$ must be chosen
 of order $100 - 200$ GeV. The resulting physical Higg boson mass
is near 150 GeV, and the $SU(2)$ singlet scalar obtains a mass
of a few 100 GeV (for details see Ref.~\cite{BS}).

\section{Conclusions}

We have constructed an example of a two-stage little Higgs: A little Higgs theory
which is embedded in a linear sigma model whose scalar fields are the pseudo
Nambu-Goldstone bosons  of a UV little Higgs theory themselves.
From the bottom up our theory is the 
Standard Model embedded in an $SU(3)$ little Higgs theory. The $SU(3)$ 
symmetry breaking is due to vevs for two triplet scalar fields which are littlest Higgses
of two UV $SU(7)/SO(7)$ sigma models. The theory is weakly coupled
all the way to its UV cutoff at $(4\pi)^3 M_W \sim 100$~TeV. 

Ordinary little Higgs theories have a cutoff of 10 TeV. Higher dimensional
operators suppressed by this scale which break approximate symmetries
of the Standard Model are very tightly constrained and must have small
coefficients. It is therefore 
interesting to see if UV extensions exist which predict (or at least are 
consistent with) small coefficients for these operators.
Experimental signatures which probe such high scales include proton decay,
lepton flavor violation, flavor changing neutral currents (FCNCs), $K-\bar K$
mixing and CP violation.

We find that the situation is very interesting, some of the constraints are
non-trivial and yield experimental signatures:

Rare processes such as $\mu\rightarrow e \gamma$ probe high energies.
Since the masses of charged leptons and their heavy $SU(3)$ partners
(heavy neutrinos) need not be aligned, we expect lepton flavor violation from
heavy neutrino loops even in the absence of SM neutrino masses.
Generic heavy neutrino masses are already ruled out by
$\mu\rightarrow e \gamma$ and the experimental constraint
requires either degeneracy of the heavy neutrinos at the
$\sim$1\% level or else mass alignment.
Note that there are no tree level FCNCs in the lepton sector
because the three generations are treated universally. 

On the other hand, we do expect
tree level FCNCs in the quark sector because the $Z'$ couplings to the third
generation of quarks are different from those to the first and second generation. 
We estimate that the contribution to $b\rightarrow s l^+ l^-$  from the $Z'$ is
comparable to the standard model contribution which makes this an interesting
signature. Note that this signature arises in all models with $SU(3)$ extensions 
of the $SU(2)_L$ gauge group where anomalies cancel between different 
generations.  

Furthermore, box diagrams with heavy partners of the quarks and
gauge bosons in the loop contribute to $K-\bar K$ mixing. Arbitrary
quark partner masses and mixing angles are already ruled out. These 
constraints - which become even more stringent in the presence
of new phases and CP violation - can be avoided if the
heavy quarks are approximately  degenerate. The situation is similar to
the minimal supersymmetric Standard Model~\cite{MSSM} where the box diagram with
superpartners requires flavor universal squark mass matrices.
It is an interesting challenge to build models which include a mechanism
that predicts degenerate heavy quarks.

Baryon number is an accidental symmetry of our theory so that proton decay
 is not predicted. Protons may still decay if there is baryon number violation
 above the cutoff. 
 
Signatures for the LHC include heavy $SU(3)$ gauge boson
and fermion partner production and decays. The upstairs little
Higgs theory can only be probed indirectly or with a higher energy collider
and could be distinguishable from other little Higgs phenomenology~\cite{pheno1,pheno2}.

\section*{Acknowledgements}

M.S. and W.S. thank the theory group at Johns Hopkins University for its hospitality during
early stages of this project. We also thank the Aspen Center for Physics for
its hospitality during intermediate stages of the project, and M.S. thanks the theory group
at Yale for its hospitality during completion of the project.
D.K. is supported by DOE OJI and NSF grants  P420D3620414350 and P420D3620434350.
M.S. is supported by DOE OJI,  the DOE grant DE-FG02-90ER-40560, 
and a grant from the SLOAN foundation.
W.S. is supported by DOE OJI and by the DOE grant DE-FG02-92ER-40704.

\end{document}